\documentclass[twocolumn,showpacs,showkeys,aip,jcp,reprint]{revtex4-1}
\usepackage[latin1]{inputenc}
\usepackage{amsmath}
\usepackage{amsfonts}
\usepackage{amssymb}
\usepackage{graphicx}
\usepackage{subfigure}

\begin{document}

\title{Wave Packet Simulations of Antiproton Scattering on Molecular Hydrogen}

  \author{Henrik Stegeby}
  \affiliation{Department of Chemistry - {\AA}ngstr\"om, Uppsala University Box 518, 75120 Uppsala, Sweden}
  \author{Markus Kowalewski}
  \affiliation{Department of Information Technology, Uppsala University Box 337, 75105 Uppsala, Sweden}
  \email{mkowalew@uci.edu}
  \altaffiliation{Now at: Chemistry Department, University of California, Irvine, California 92697-2025, United States}
  \author{Konrad Piszczatowski}
  \affiliation{Department of Chemistry - {\AA}ngstr\"om, Uppsala University Box 518, 75120 Uppsala, Sweden}
  \author{Hans O. Karlsson}
  \email{hans.karlsson@kemi.uu.se}
  \affiliation{Department of Chemistry - {\AA}ngstr\"om, Uppsala University Box 518, 75120 Uppsala, Sweden}


\date{\today}
\begin{abstract}
The problem of antiproton scattering on the molecular hydrogen  is investigated by means of wave packet dynamics. The electronic potential energy surfaces of the antiproton ${\rm H}_2$ system are presented within this work. Excitation and dissociation probabilities of the molecular hydrogen for collision energies in the ultra low energy regime below 10\,eV are computed.
\end{abstract}


\maketitle

\section{Introduction}
The antimatter-matter asymmetry is an active area of physics adressing fundamental questions concerning the evolution of the universe, the validity of the CPT symmetry and gravitational differences between matter and antimatter \cite{alpha:13,Villata:11}. 

Even though our universe seems to be solely composed of regular matter it is possible to create and study antimatter in the laboratory \cite{Baur:96,Christian:97}.
The recent advances in the field of antimatter include the production and trapping of
antihydrogen \cite{Madsen10} as well as cooling to a stable electronic state \cite{alpha:11}. Antihydrogen is one of the simplest examples of an atomic system
composed of antimatter. However, to increase the understanding of antimatter systems the interaction between matter and antimatter is of special interest.
Mixed matter-antimatter many-body systems provide a field of study, which exceeds the pure
annihilation effects and allows for comparison with the dynamics of ordinary molecular systems.
Examples of such many body systems includes, e.g., antihydrogen-hydrogen \cite{Stegeby:11}, antiprotonic helium \cite{ASACUSA:11b,Hiroyuki:07} and various antiproton - gas collisions investigated by the ASACUSA collaboration in the antiproton collisional kinetic energy regime of 2-11 keV \cite{ASACUSA:10}.

In this work we have investigated the collision  an antiproton ($\bar{p}$) with  molecular hydrogen as a  model system for more complex antiparticle - molecular interactions.
The $\rm H_2$-$\bar{p}$ system is interesting from a practical point of view as residual ${\rm H_2}$ gas exists in most high energy colliders, and antiprotons are created in various experiments, resulting in scattering of antiprotons on ${\rm H_2}$ molecules \cite{Hiroyuki:07}.
Previous work on this system focused on  the high-energy \cite{Pindzola:10} and the low-energy collision region investigating  antiproton and proton cross sections for ionization and excitation of hydrogen molecules as well as energy spectra of the ionized electrons\cite{Luehr:08,Luehr:09,ASACUSA:10,Abdurakhmanov13}.

In previous work the influence of collisions in the energy regime above 100\,eV was investigated which mainly leads to ionized reaction products \cite{Agnello95prl}.
However, the ultra low energy regime below 10\,eV \cite{Trujillo11} opens up a new regime of matter-antimatter chemistry, since this energy is comparable to the chemical bond energy. In this region the molecular hydrogen is vibrationally excited, but remains in its electronic ground state.
The antiproton behaves like another nucleus in the molecular system, whereas  its negative charge produces an interaction potential, which is on the order of magnitude of an electronic state in an atom.

In this paper we have computed potential energy surfaces (PES) for the lowest electronic states of the $\rm H_2$-$\rm\bar{p}$ system and performed wave packet simulations of the collision on the electronic ground state.
The wave packet simulation provide information about probability distributions of the vibrational states of the ${\rm H_2}$ molecule after the collision, as well as the dissociation probability.
The paper is organized as follows: In section \ref{sec:model} the quantum dynamical model is introduced. In section \ref{sec:results} the potential energy surfaces (PES) of the $\rm H_2$-$\rm \bar{p}$ system and the results of the wave packet calculations are presented followed by a discussion in section \ref{sec:discussion}.

\section{Quantum Dynamics Model}
\label{sec:model}
\begin{figure}
\centering
\includegraphics[width=.6\columnwidth]{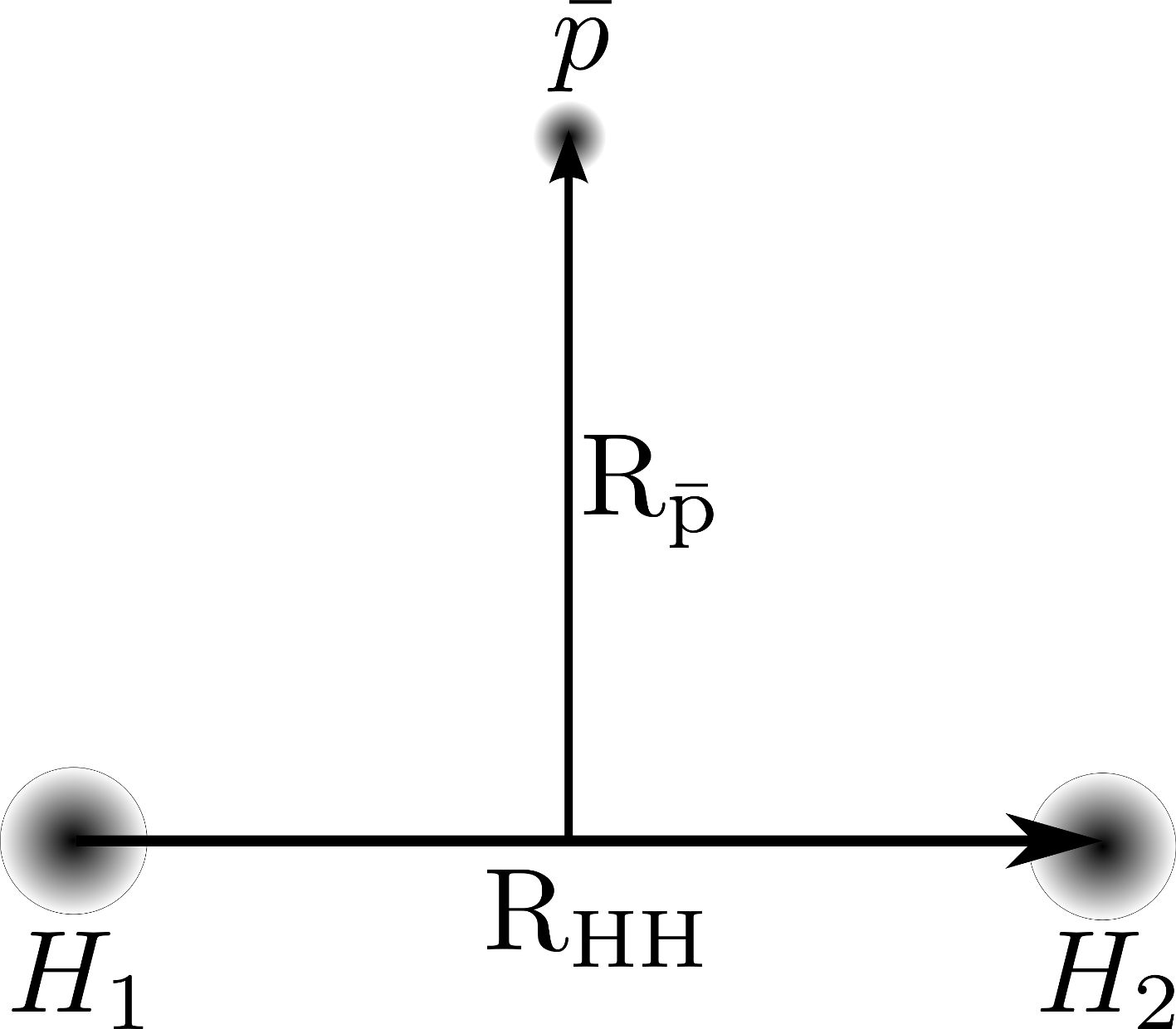}
\caption{Jacobi coordinates for the $\rm H_2 - \bar{p}$ system in a T-shaped configuration. The internuclear separation between the hydrogen atoms $\rm H_1$ and $\rm H_2$ are varied along the $R_{HH}$ axis and the distance from the barycenter of the two hydrogen atoms to the antiproton is the $R_{\bar{p}}$ axis.}
\label{systempic}
\end{figure}
We have considered a T-shaped configuration  of the $\rm H_2$-$\bar{p}$ system corresponding to a zero impact parameter, as illustrated  in Fig. \ref{systempic}, which is expected to have a maximum interaction between the antiproton and the vibrational degree of freedom of the molecular hydrogen.
Jacobi coordinates are used  to describe the inter-atomic distances between the H-atoms ($R_{HH}$) and the antiproton distance to the center of mass of $\rm H_2$ ($R_{\bar{p}}$). The angle of the H-H bond axis and $R_{\bar{p}}$ is fixed to be $90$ degrees. 
This choice of coordinates focuses specifically on the vibrational and dissociation dynamics of $\rm H_2$ in collision event with $\bar{p}$. Other reaction pathways like the protonium formation \cite{Trujillo11} are neglected in this model.

To simulate the collision between the hydrogen molecule and the antiproton we solve the time-dependent Schr\"odinger equation for the nuclei
\begin{equation}
\label{eq:TDSE}
i\hbar\dfrac{\partial}{\partial t} \Psi(R_{HH},R_{\bar{p}},t)=\hat H \Psi(R_{HH},R_{\bar{p}},t)\,,
\end{equation}
numerically. Atomic units are used (i.e. $\hbar=m_e=e=4\pi\epsilon_0=1$) in what follows.
The Hamiltonian in Jacobi coordinates reads:
\begin{align}
\label{eq:H}
\hat{H}&=\hat{T}_{R_{\bar{p}}} + \hat{T}_{R_{HH}} + \hat{V}({R_{HH},R_{\bar{p}}}) =\\
&- \frac{1}{2\mu_{HH}} \frac{\partial^2}{\partial R_{HH}^2} - \frac{1}{2\mu_{{\bar{p}}}} \frac{\partial^2}{\partial R_{\bar{p}}^2} + \hat{V}(R_{HH},R_{\bar{p}})\nonumber \,,
\end{align}
where $R_{HH}$, and $R_{\bar{p}}$ 
are the Jacobi coordinates as defined in Fig. \ref{systempic}.
The reduced masses and $\mu_{{\bar{p}}}$ and $\mu_{{HH}}$
are given by
\begin{eqnarray}
\mu_{HH} &=& \dfrac{m_p}{2}\\
\mu_{\bar{p}} &=& \dfrac{2m_p^2}{(m_p + 2m_p)} = \dfrac{2}{3}m_p \,,
\end{eqnarray}
where $m_p$ is the rest mass of the proton. The choice of the Jacobi coordinates
ensures that the system is in the center of mass frame. 
The molecular potential energy surface (PES) $\hat V$ represents the adiabatic electronic ground state potential of the system and is obtained by solving the electronic Schr\"odinger equation under the assumption that the Born-Oppenheimer approximation is valid.
The collision of the molecular hydrogen with the antiproton in the given coordinate system $R_{HH}$, $R_{\bar{p}}$ is simulated by solving eq. \ref{eq:TDSE} numerically with
a Chebyshev time propagation scheme \cite{Ezer:84}. 
A perfectly matched layer (PML) \cite{Nissen:10} at the edges of the numerical grid has been used absorb the wavefunction and to avoid spurious reflections.

The initial condition is chosen to meet the given physical conditions of an antiproton approaching a hydrogen molecule. The initial wave function can be written as a product state
of the $\rm H_2$ vibrational ground state wave function $\phi_0$ and a Gaussian wave packet representing the antiproton:
\begin{align}
\label{eq:Psi0}
\Psi(R_{HH}, &R_{\bar{p}}, t=0) = \\
&\sqrt{\frac{2}{\sigma \sqrt{2\pi}}} {\rm e}^{ik R_{\bar{p}}-(R_{\bar{p}}-R_0)^2/2\sigma^2} \phi_0(R_{HH})\,.\nonumber
\end{align}
The collision energy $\rho_0={p^2}/{2\mu_{\bar p}}$ enters eq. \ref{eq:Psi0}
through the momentum $p=\hbar k$. The width $\sigma$ of the wave packet is assumed to be 0.3 \AA, which corresponds to a width of 35\,meV in kinetic energy. In the limit of
$R_0\rightarrow\infty$ i.e. of a separated reactants the PES becomes flat and the Schr\"odinger equation becomes separable in the molecular coordinates justifying the choice
of the wave-packet, eq. \ref{eq:Psi0}. A value of $R_0 = 10\,{\rm \AA}$ is chosen as a good approximation on a finite surface used in the numerical model.

By describing the system with the non-relavistic Schr\"odinger equation, the particle anti-particle annihilation of the proton and the antiproton is neglected. 
However, a direct p-$\rm \bar p$ collision has a low probability, specially
for the given setting (impact parameter of zero).
This is justified by the wave packet simulations which show
that the probability density of the wave function is low at the positions representing a direct p-$\rm \bar p$ contact.
A comparison of the simulation time scale for a single collision event with experimentally determined annihilation time constants \cite{Agnello95prl} shows that the annihilation probability is indeed low and annihilation effects are expected to be small and to have no significant influence here.

\section{Results}
\label{sec:results}
\subsection{Potential energy surfaces}
\begin{figure*}
\centering
\subfigure[]{\includegraphics[height=7cm,width=5cm]{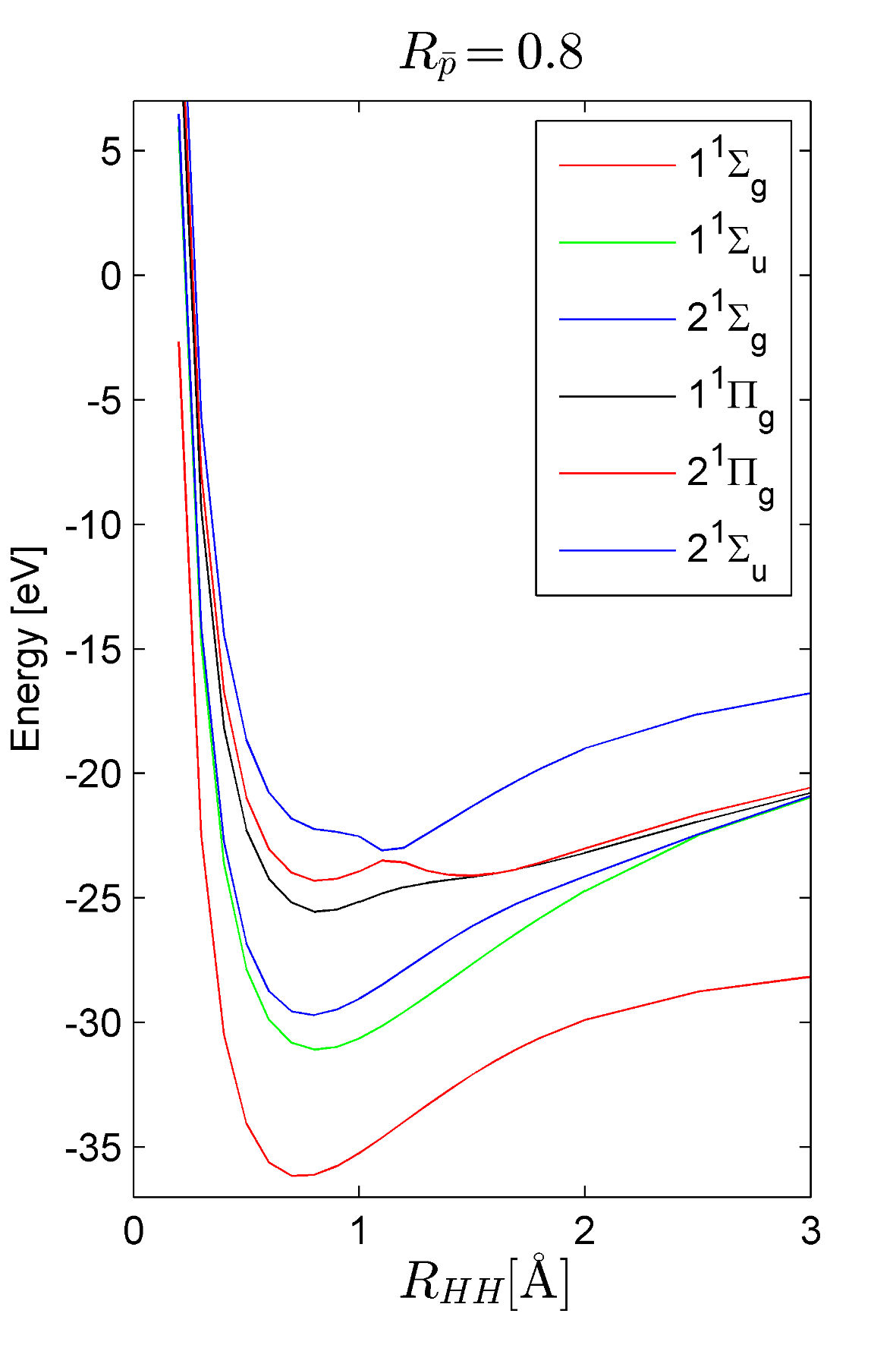}}
\subfigure[]{\includegraphics[height=7cm,width=5cm]{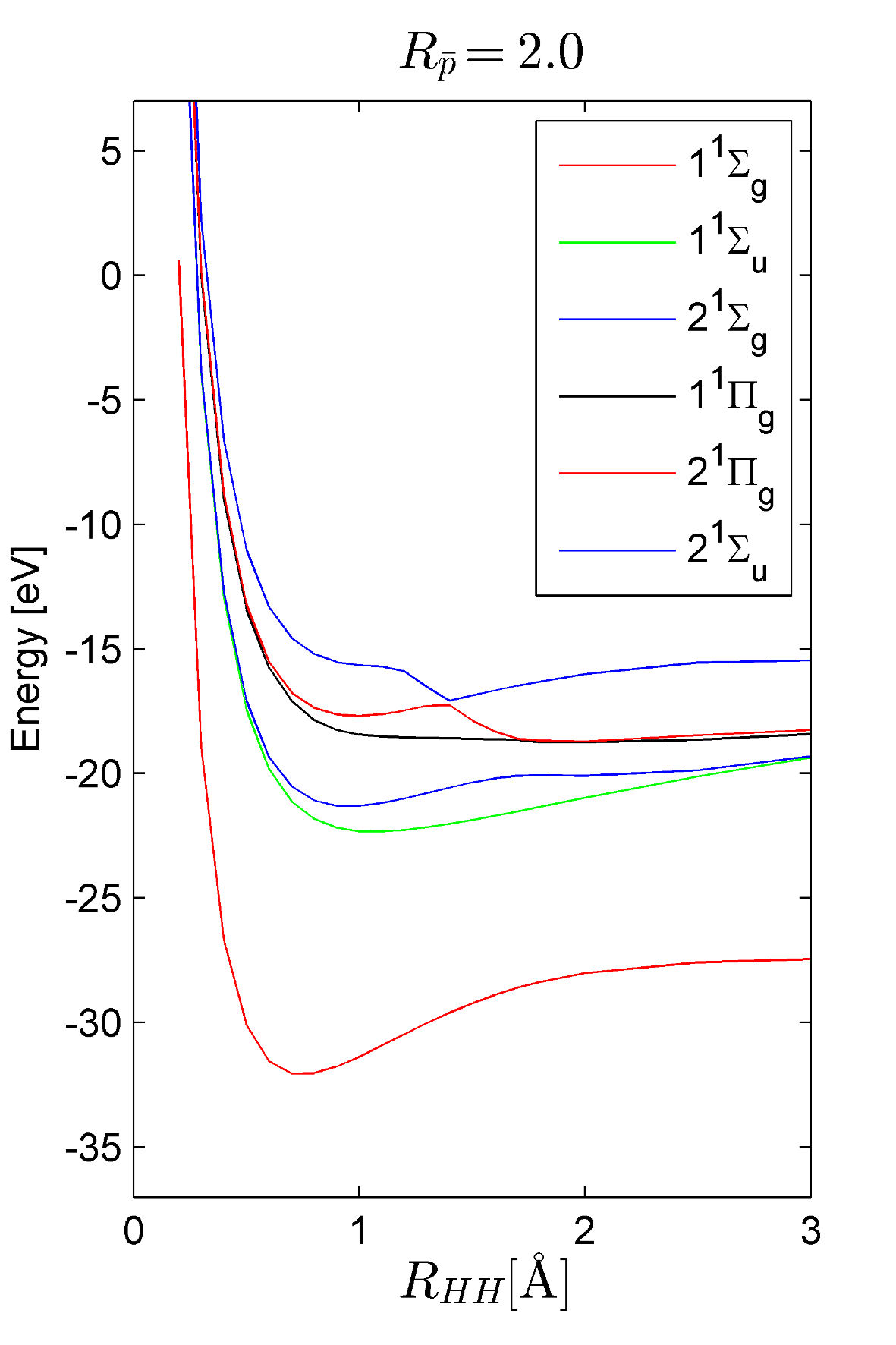}}
\subfigure[]{\includegraphics[height=7cm,width=5cm]{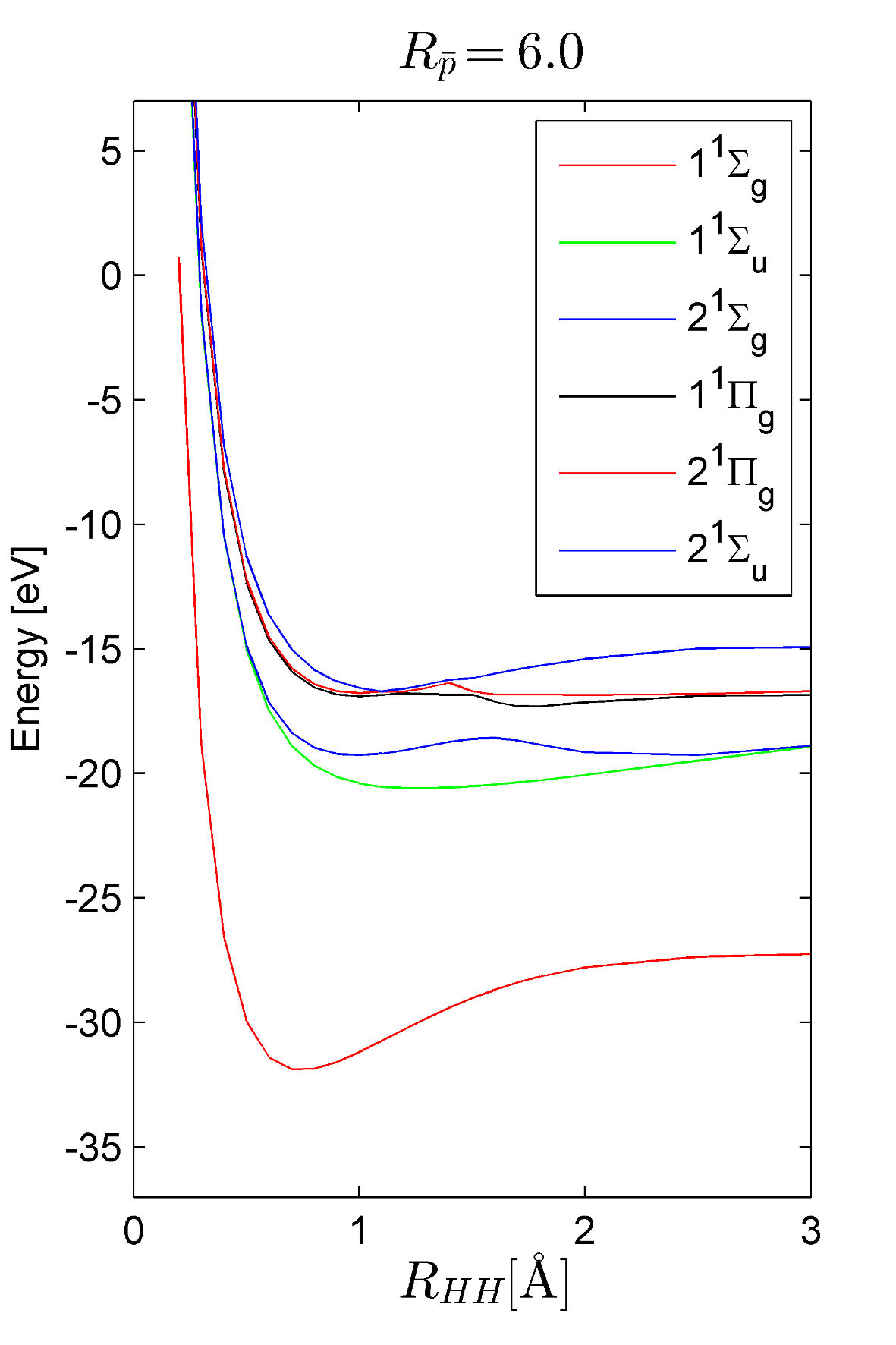}}
\caption{Slices through the PES for different values of $\rm R_{\bar{p}}$. (a) 0.8\,\AA, (b) 2.0 \,\AA, and (c) 6.0\,\AA. The ground state ($\rm 1^1\Sigma_g$) and the first - to fifth excited states ($\rm H_2$ state symmetries $\rm 1^1\Sigma_u$, $\rm 2^1\Sigma_g$, $1^1\Pi_g$, $2^1\Pi_g$ and $\rm 2^1\Sigma_g$) energy are varied along the internuclear separation between the hydrogen atoms $\rm R_{HH}$.
The labels of the electronic states refer to the electronic states of unperturbed $\rm H_2$ in the $\rm D_{\infty h}$
point group.
A potential energy of 0\,eV refers to the limit of separated particles.}
\label{fig:PES}
\end{figure*}

The electronic ground state and the first five excited states potential energy surfaces have been
calculated with the full configuration interaction (FCI) method using MOLCAS program package \cite{MOLCAS}. The Atomic Natural Orbital basis set of valence triple zeta accuracy with polarization functions (ANO-L-VTZP) basis set  \cite{MOLCAS} is used to describe the two electrons. The influence of the antiproton on the electronic states of the $\rm H_2$ molecule is simulated by adding a negative point charge to the geometry definition.
A comparison between the ground state and the first few excited electronic state potential energy surfaces (PES) is presented in one-dimensional slices at different antiproton distances in Fig. \ref{fig:PES}.
In Fig. \ref{fig:PES} one can see that the electronic ground state is well separated from the excited state surfaces and no avoided crossings or conical intersections between the ground state and an excited state occurs.
However, curve crossings can be observed between the $1^1\Pi_g$ and the $2^1\Pi_g$ state at $R_{HH} \approx 1.6\,\rm\AA$ and  $2^1\Pi_g$ and the $2^1\Sigma_g$ state
at $R_{HH} \approx 1.2$\,\AA.
Note that for the hydrogen molecule only avoided crossings can occur, whereas the triatomic  H$_2$-$\bar{p}$ system can have  conical
intersections \cite{Domcke,Domcke12arpc} between states of the same symmetry.
As the antiproton approaches the $\rm H_2$ molecule the PES energy and the difference between the vertical excitation energy and the groundstate energy decreases, while the $1^1\Pi_g$, $2^1\Pi_g$ and $\rm 2^1\Sigma_g$ states increase their separation in energy.  For the $\rm 2^1\Sigma_g$ and $1^1\Pi_g$ excited state crossings have been observed. The groundstate energy curve becomes identical to the ground state curve for unperturbed $\rm H_2$  as $R_{\bar{p}} \rightarrow \infty$. The dissociative limit of the $\rm H_2$ groundstate approaches -27.21\,eV, which corresponds to the sum of two H atoms in their groundstate, as $R_{HH} \rightarrow \infty$. 

\begin{figure}
\centering
\includegraphics[width=\columnwidth]{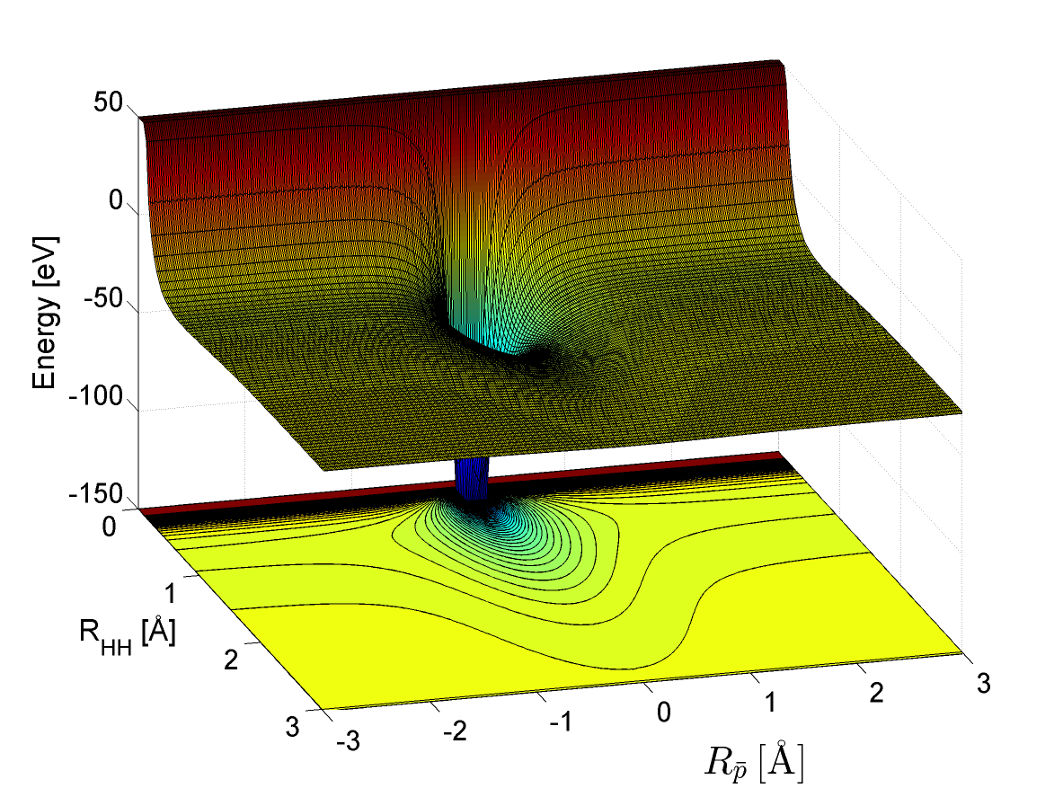}
\caption{PES for the $\rm H_2 - \bar{p}$ systems ground state, calculated in dependence of the $\rm H_2$ separation and the antiproton distance.
For clarity the potential is truncated above +50\,eV. Note that the potential converges to $-\infty$ at $(0,0)$.}
\label{molcaspotsurf}
\end{figure}

A two-dimensional plot of the electronic ground state PES in the T-shaped configuration, is shown in Fig. \ref{molcaspotsurf}. A deep potential well is found where the antiproton passes through the axis of the molecular hydrogen
caused by an attractive Coulomb interaction between the protons and the antiproton.
The theoretical value in the origin of the coordinate system $\{R_{HH},R_{\bar p}\}$ can be directly derived from the Coulomb interactions $V_C$ by taking the limit  $\{R_{HH},R_{\bar p}\} \rightarrow 0^+$ for the interaction potential:

\begin{align}
\label{eq:Vc}
\lim_{\{R_{HH} ,\, R_{\bar{p}}\} \rightarrow  0^+} V_C &=
\lim_{\{ r_{1 \bar{p}} ,\,  r_{2 \bar{p}} ,\, r_{12} \} \rightarrow  0^+}\frac{e_p e_{\bar{p}}}{r_{1\bar p}} + \frac{e_p e_{\bar{p}}}{r_{2\bar p}} + \frac{e_{p}^2}{r_{12}}\\
&=\lim_{ \{ R_{HH} ,\,  R_{\bar{p}} \} \rightarrow  0^+} \frac{1}{R_{HH}} - \frac{2}{\sqrt{R_{\bar{p}}^2 + (R_{HH}/2)^2}} = -\infty \,,\nonumber
\end{align}
where $r_{1\bar p}$, $r_{2\bar p}$ and $r_{12}$ are the distances between proton 1 and the antiproton, proton 2 and the antiproton, and the distance between proton 1 and 2 respectively. Here $e_p=+1$ and $e_{\bar p}=-1$ are the charge
of the proton and the antiproton respectively. From eq. \ref{eq:Vc} it becomes clear that the limit diverges to minus infinity. This is indicated in Fig. \ref{molcaspotsurf} showing a potential hole diverging to minus $-\infty$.

\subsection{Quantum Dynamics Simulations}
\label{section:simulations}
The H$_2$ antiproton collision is simulated by solving eq. \ref{eq:TDSE}
numerically using the  Chebyshev time propagation scheme \cite{Ezer:84}, combined  with the Fourier method \cite{Tannor}, for 
different collision energies.
The initial conditions are constructed from eq. \ref{eq:Psi0} and cover collisions with an incoming energy between 0.01 and 10 eV.
The PES which enters through the Hamiltonian (eq. \ref{eq:H}) is the $\mathrm{1^1\Sigma_1}$ electronic ground state as it is shown in Fig. \ref{molcaspotsurf}. We assume that the electronic ground state of the H$_2$-$\bar p$ system is sufficient to describe the collision in the energy
regime considered. This is justified by the absence of curve crossings
between the ground state and the excited states.

\begin{figure*}
\center
\subfigure{
\includegraphics[width=0.3\textwidth]{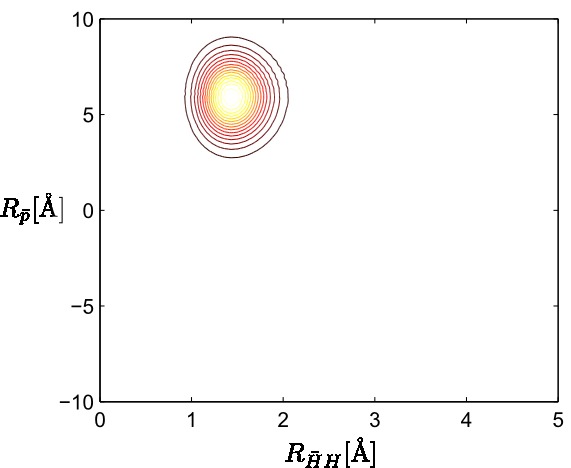}
\includegraphics[width=0.3\textwidth]{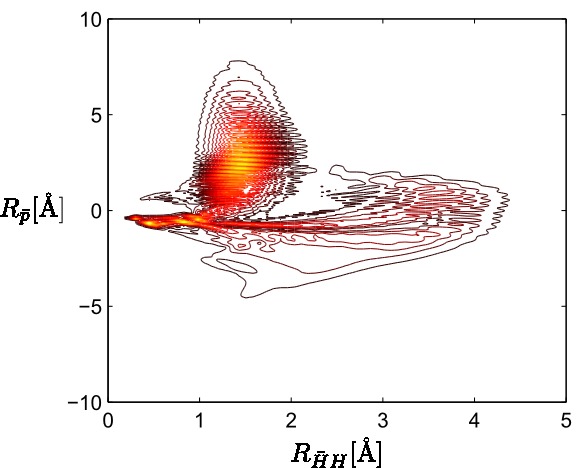}
\includegraphics[width=0.3\textwidth]{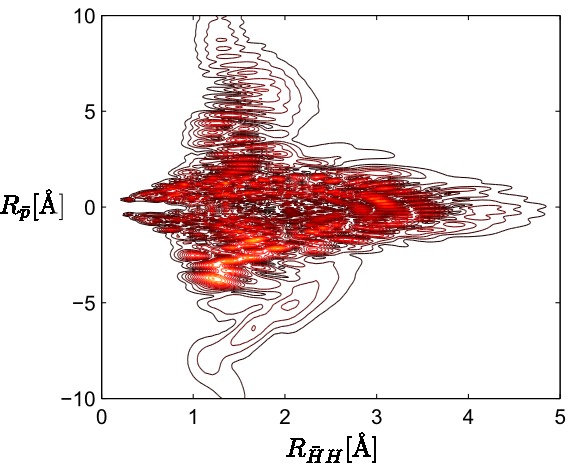}
}
\caption{Time evolution of the antiproton wave packet with a kinetic energy of $\rho_0=3$\,eV at times $t=60$ fs,  $t=110$ fs and $t=160$ fs (from left to right).\label{timeev}}
\end{figure*} 
The time evolution of a wave packet with an initial kinetic energy of 3\,eV
is shown in Fig. \ref{timeev} for 3 different times. As the wave packet approaches
the deep potential well the strong attraction of the Coulomb force becomes visible (Fig. \ref{timeev}b)).
When the antiproton transverses the potential singularity, the wave function is either reflected, transmitted, or deflected towards the H$_2$ dissociation. 
As the potential is a conservative potential in this model the wave packet can not be trapped in the location around the potential singularity.
Eventually it will be scattered into separate fragments.
The quantitative results of the inelastic scattering event will be presented in the following
sections. 

\subsubsection{$\rm H_2$ vibrational excitation}
\label{sec:exc}
The vibrational excitation after the collision event is analyzed separately for the reflected and the transmitted part of the wave function.
In the asymptotic limit of a large $|R_{\bar{p}}|$
the Schr\"odinger equation for the Hamiltonian (eq. \ref{eq:H}) becomes separable and the vibrational excitation can be expressed in terms of vibrational eigenfunctions $\phi_{\nu}(R_{HH})$ of $\rm H_2$.
The probability of finding the system in a  vibrational eigenstate $\phi_\nu$ at a certain time $t$
is given by the projection against the total nuclear wave function $\Psi(R_{HH},R_{\bar{p}^0},t)$:
\begin{equation}
\label{eq:cnu}
|c_\nu(t)|^2= \int \lvert \phi_\nu(R_{HH}) \Psi(R_{HH},R_{\bar{p}^0},t) \rvert^2 dR_{HH}\,.
\end{equation}
The distance from the H$_2$ center is chosen to be $R_{\bar{p}^0}=10$ \,\AA,  which is far enough from the interaction region such that the separability assumption is sufficiently fulfilled.
The total probability $P_\nu$
is then obtained by integrating eq. \ref{eq:cnu} over time \cite{Zamith:99}:
\begin{equation}
 P_\nu=\frac{1}{N^{\tau}}\int_{0}^{\tau}| c_\nu(t)|^2 dt\,.
\end{equation}
The time end point $\tau$ (500 fs) is chosen such that less than 1\% of the wave packet remains after absorption at the PML. A time step 242 as has been chosen and is sufficiently short to allow for an accurate numerical integration over time.
 
The antiproton is scattered in all directions and integral bounds have to set
such that the reflected part of the wave function $\Psi_R$ and the transmitted part of the wave function $\Psi_T$ is captured. The sum over the coefficients  corresponding to the these probability distributions is the normalization coefficient $N^{\tau}$ we use, i.e. 
\begin{equation}
N^{\tau} = \sum_{\nu} \left( \int_{t=0}^{\tau}  |c^R_\nu(t)|^2 dt + \int_{t=0}^{\tau} | c^T_\nu(t) |^2 dt \right)\,.
\label{norm}
\end{equation}
The normalization $N^{\tau}$ is chosen such that the sum of the probabilities for the reflected and transmitted part is unity. 

The total probability for the elastic scattering $P^E(\nu=0)$ is given by
 \begin{equation}
 P^E_{\nu=0}=\frac{ \int_{t=0}^{\tau}  |c^R_0(t)|^2 dt + \int_{t=0}^{\tau} | c^T_0(t) |^2 dt}{N^{\tau}}\,, 
 \end{equation}
which, together with the dissociation probability $P_{diss}$ described in section \ref{sec:dissociation} gives the resulting inelastic scattering probability.

\begin{figure}
\center
\subfigure[]{
\includegraphics[height=6.5cm,width=7.5cm]{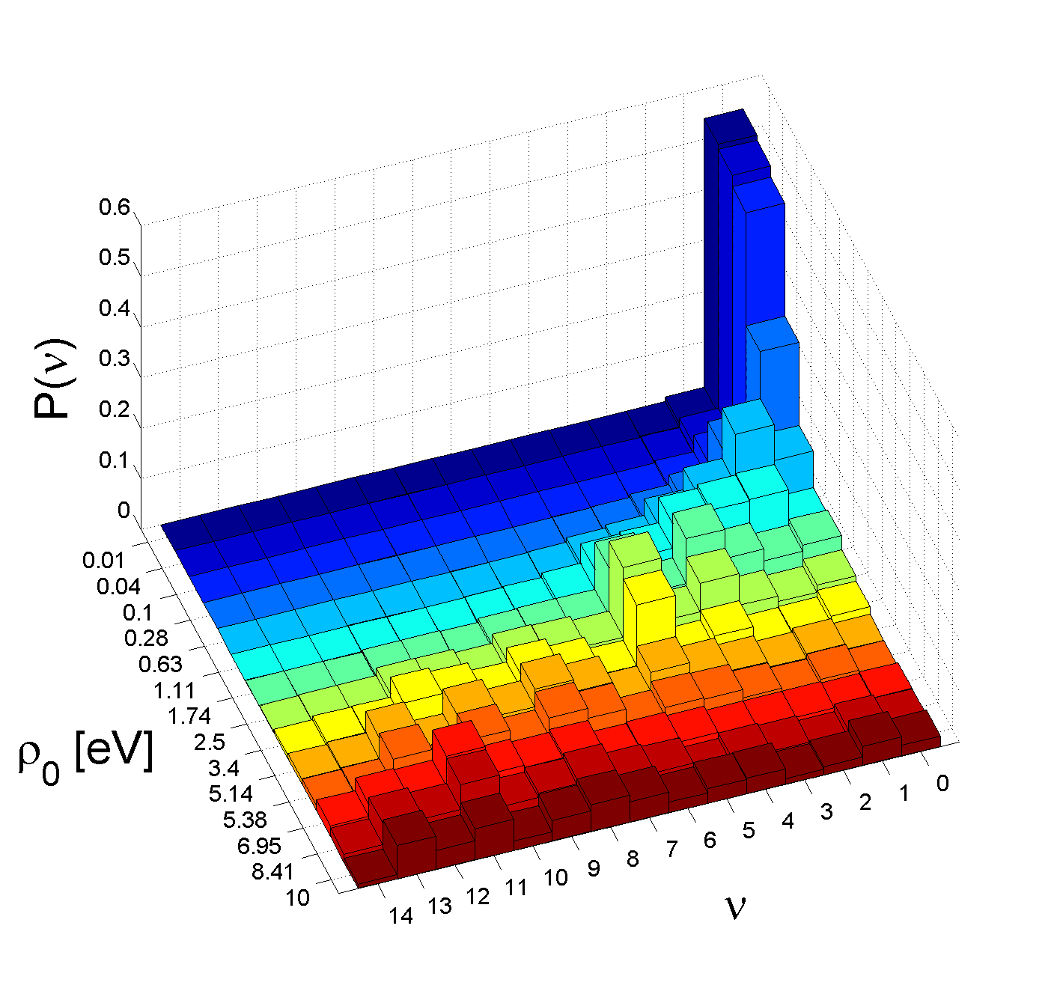}
}
\subfigure[]{
\includegraphics[height=6.5cm,width=7.5cm]{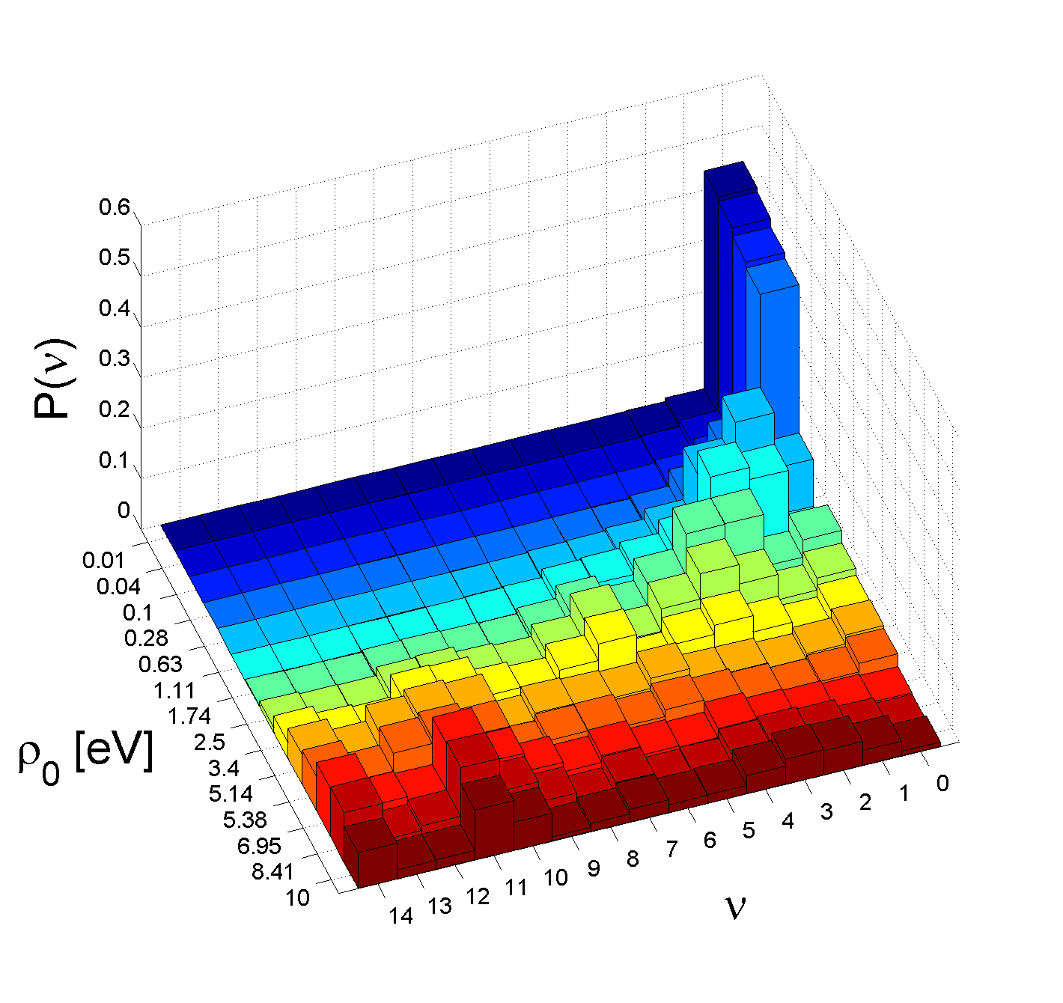} 
}
\caption{The probability distribution for the reflected (a) and the transmitted (b) wavefunction for the $\rm H_2$ molecule to end up in vibrational excitation state $\nu$=0 to 14 versus initial antiproton kinetic energy $\rho_0$. The probabilities of both diagrams sum up to one for a specific energy value.\label{ProbDistFig}}
\end{figure}
\begin{figure}
\centering
\includegraphics[width=\columnwidth]{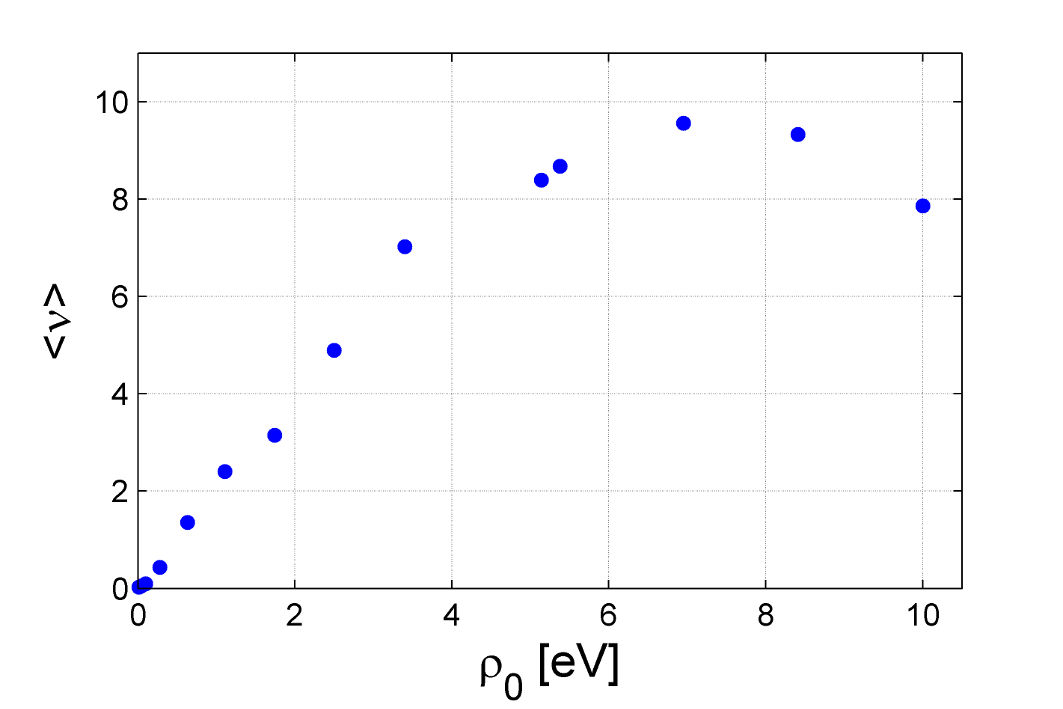}
\caption{The mean vibrational quantum number $\langle \nu \rangle $ distribution versus initial antiproton kinetic energy $\rho_0$ for the sum of the reflected and the transmitted wavefunction.}
\label{fig:vmean}
\end{figure}

The vibrational distributions for the transmitted and reflected part are shown in Fig. \ref{ProbDistFig}. Until 0.1\,eV elastic scattering dominates the process
and $\rm H_2$ remains in its vibrational ground state.
At 0.28\,eV the reflected part begins to slightly broaden its vibrational distribution
while the transmitted part still remains in $\nu=0$.
This is the main difference between the reflected and the transmitted vibrational distributions, which otherwise are very similar.
As the kinetic energy is increased further, $\nu$ displays a broad distribution of the
vibrational states.
For comparison the mean vibrational quantum numbers
\begin{equation}
\langle \nu \rangle = \sum_{\nu=0}^{14} \nu P_\nu
\end{equation}
of the H$_2$ vibrations after the collision are shown
in Fig. \ref{fig:vmean}. Due to the small difference between the reflected and transmitted parts only the sum of both is shown. 
The mean vibrational quantum number increases linearly until the $\rm H_2$ dissociation energy of 4.52\,eV and peaks around 7\,eV with a maximum around $\langle\nu\rangle=10$.
The mean vibrational quantum number can be regarded as a good measure until about 6\,eV
where the dissociation probability reaches a threshold and increases.

\subsubsection{$\rm H_2$ Dissociation}
\label{sec:dissociation}
The dissociation probability of H$_2$ is analyzed by using the
probability flux derived from the continuity equation \cite{Tannoudji}:
\begin{equation}
 \label{eq:Cont}
 \dfrac{\partial\Psi^*\Psi}{\partial t} + \nabla \cdot \vec{j} = 0 \,,
\end{equation}
where $\vec{j}$ is the probability flux. Eventually, the probability flux $F(t)$
reads:
\begin{equation}
\label{eq:F}
F(t)=-\dfrac{2}{\hbar} \int_V \Im\left( \Psi(R_{HH}, R_{\bar p},t)^* \hat T \Psi(R_{HH}, R_{\bar p},t) \right) dV\,,
\end{equation}
where $\Im$ denotes the imaginary part, and $V$ is the volume corresponding to a bound H$_2$ in the coordinate system $R_{HH}$, $R_{\bar p}$. The operator $\hat T$
is the kinetic operator from the Hamiltonian (eq. \ref{eq:H}).
By using a step function $\Theta(R_{HH}-R_{HH,0})$ which drops to zero at $R_{HH}=15 $\,\AA, the probability flux describing the dissociation of H$_2$ is obtained:
\begin{align}
F_{diss}(t)=&-\dfrac{2}{\hbar} \int_{-\infty}^{\infty} \Theta(R_{HH}-R_{HH,0})\\
&\times \Im\left( \Psi(R_{HH}, R_{\bar p},t)^* \hat T \Psi(R_{HH}, R_{\bar p},t) \right) d\mathbf{R}\,. \nonumber
\end{align}
The integration over time yields the dissociation probability:
\begin{equation}
P_{diss} = \int_0^T F_{diss}(t) dt
\end{equation}
\begin{figure}
\centering
\includegraphics[width=\columnwidth]{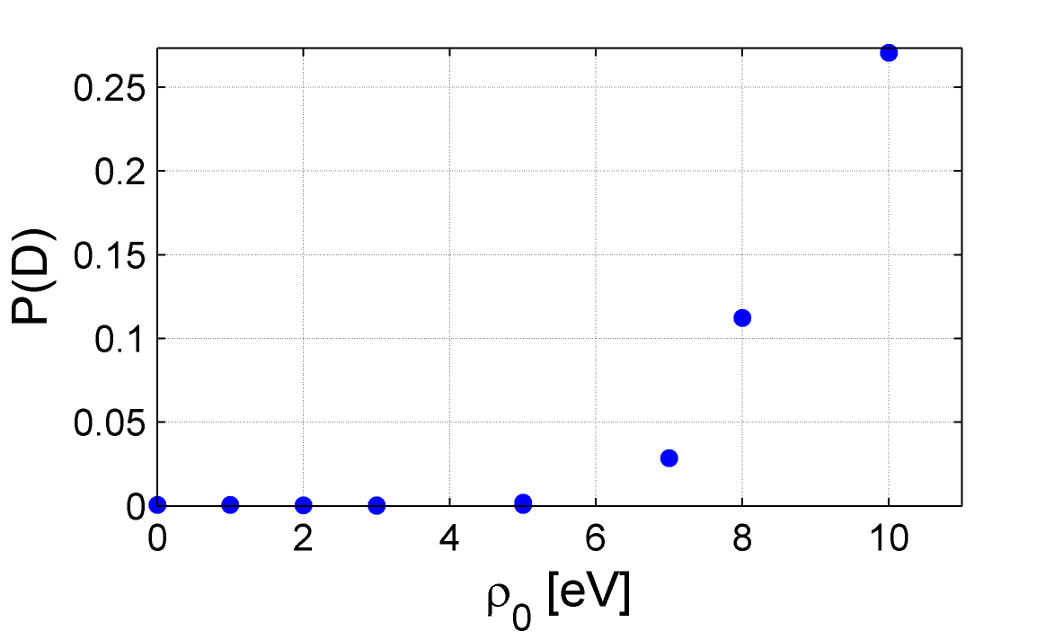} 
\caption{Dissociation probability P(D) of the $\rm H_2$ molecule versus initial antiproton kinetic energy $\rho_0$, displaying a rapidly growing behavior after 6 eV.}\label{dissociationfig}
\end{figure}
The dissociation probability is shown in Fig. \ref{dissociationfig}. 
It is close to zero until it reaches a dissociation threshold at 6\,eV.
Between 7 and 10\,eV the dissociation probability increase linearly to
25\,\%. Note that the dissociation energy of $\rm H_2$ is 4.52\,eV.

\section{Discussion}
\label{sec:discussion}
The ground state of the $\rm H_2$-$\bar{p}$ system is separated by more than 5\,eV from the first excited state, which can be seen from Fig. \ref{fig:PES}, and thus the effect of non-adiabatic couplings can be neglected if the $\rm H_2$ is initially in its electronic ground state. 
We notice, however, that the excited states $2^1\Pi_g$ and $2^1\Sigma_u$ of the coupled system show  clearly a region of curve crossing (Fig. \ref{fig:PES}(b)), which needs to be taken in consideration for higher collision energies.
In a three body system an avoided crossing can become a conical intersection (CI) \cite{Domcke}
with a truly degenerate point. The third degree of freedom necessary to form a CI originates from an anti matter
particle with a negative charge instead of an ordinary nucleus, making it a special case. This can also be viewed as a more general case of CIs. Other examples for introducing CIs through a non-standard nuclear degree of freedom are light induced CIs \cite{Demekhin13jcp}.

The ground state potential in two dimensions (Fig. \ref{molcaspotsurf}) demonstrates the unique features of the interaction with a negative charge and the hydrogen molecule.
The PES is characterized by its deep Coulomb hole diverging to minus infinity.
The form of the potential suggests that a bound vibrational ground state could exist. However, this ground state -- representing a captured antiproton --
could be subject to strong annihilation effects and short lived.
In contrast to a tri-atomic molecule the $\rm H_2$-$\bar{p}$ ground state is expected to be several keV deep \cite{Desai:60} due to the mass of the antiproton (compared to an electron). Its determination and investigation however is beyond the scope of this paper.

The analysis of the vibrational distribution of $\rm H_2$ (Fig. \ref{ProbDistFig}) shows a clear threshold in collision energy below which the reaction products are mainly in the vibrational ground state. This threshold value is between 0.28 and 0.63\,eV and is consistent with the
the first vibrational excitation of $\rm H_2$ at $0.55$\,eV \cite{Irikura07jcp}. However,
the distribution has a slight asymmetry between the reflected and transmitted part at 0.28\,eV. This indicates that the slightly longer interaction time for the reflected
part allows coupling of some of the antiproton's kinetic energy into the $\rm H_2$ mode.
For collision energies beyond 3.4\,eV the vibrational distribution becomes more spread
out, which can be explained by the anharmonicity of the $\rm H_2$: at higher collision
energy more vibrational levels become accessible.
Here the mean vibrational quantum number $\langle \nu \rangle$ (Fig. \ref{fig:vmean}) provides a more clear picture. The increase in $\langle \nu \rangle$ is close to a linear
behavior for collision energies below 4\,eV. A comparison with the energy of $\nu=7$ (2.9\,eV) and the collision energy leads to the conclusion that approximately  90\,\%
of the antiprotons kinetic energy is converted into vibrational energy. 

At a collision energy of $\approx$ 6\,eV the $\rm H_2$ begins to dissociate (Fig. \ref{dissociationfig}). However at least 8\,eV is necessary to dissociate about
10\,\% of the $\rm H_2$. Compared with the dissociation energy of 4.52 eV of $\rm H_2$
this is a remarkable excess of energy which is needed to induce the $\rm H_2$ dissociation.
This indicates that the coupling between the antiproton mode and the $\rm H_2$ mode
becomes more inefficient with increasing collision energy. Such effects have been observed in other reactive
scattering systems in the form of a dynamical barrier \cite{Kowalewski14JPCA}. A simple explanation for this phenomena could
be provided by the fact, that the interaction time between the $\rm H_2$ and the antiproton becomes shorter at higher velocities and reduces the effective
time to transfer energy between the modes.

\section{Conclusions}
To describe the dynamics of an antiproton colliding with a hydrogen molecule
the electronic PES of the ground state as well as for the five lowest excited states were first computed.
The quantum chemical calculations show that it is enough to include the ground state PES to describe the dynamics in the ultra low energy regime considered here. 
For higher collision energies it might be necessary to include non-adiabatic couplings, as indicated by  the curve crossings between the excited states. The possibilities of finding
conical intersection instead of avoided crossings, compared to bare $\rm H_2$, is an exciting
example of non-conventional conical intersections.
A thorough analysis will be conducted in a future paper.
The results of the wave packet dynamics give insight into the energy redistribution
during the collision. For energies below the dissociation threshold the scattering event leaves a clear signature in the vibrational distribution of the $\rm H_2$ system. The
vibrational excitations are in accordance with the $\rm H_2$ vibrational states. The system has
a clear threshold with respect to dissociation. The theoretical calculations show that
this threshold is $\approx$ 1.5\,eV above the dissociation limit of $\rm H_2$.

\begin{acknowledgements}
M.K. acknowledges support through the Centre of Interdisciplinary Mathematics (CIM), Uppsala University. 
H.S. acknowledges support from Roland Lindh, and want to thank Alejandro Saenz and Armin L\"{u}hr for helpful discussions concerning the $\rm H_2$ - $\rm \bar{p}$ scattering problem, as well as Mickael G. Delcey for help regarding the MOLCAS program package. The computations were performed on resources provided by SNIC through Uppsala Multidiciplinary center for Advanced Computational Science (UPPMAX) under Project snic2013/1-267
\end{acknowledgements}

\end{document}